\documentclass[english,two column, showpacs]{revtex4}
\usepackage[T1]{fontenc}
\usepackage[latin1]{inputenc}
\usepackage{graphicx}
\usepackage{amsmath}
\usepackage{amssymb}
\def\a{\alpha}
\def\b{\beta}

\def\e{\varepsilon}

\def\l{\lambda}

\def\m{\mu}

\def\t{\tau}

\def\o{\omega}
\def\r{\rho}
\def\s{\sigma}

\def\S{\Sigma}

\def\D{\Delta}

\def\pd{\partial}

\def\br{{\bf r}}

\def\bn{{\bf n}}

\def\be{\begin{equation}}
\def\ee{\end{equation}}
\def\bea{\begin{eqnarray}}
\def\eea{\end{eqnarray}}
\def\nn{\nonumber}
\def\lb{\label}

\begin{document}

\title{Transport Processes in Metal-Insulator Granular Layers}
\author{Y. G. Pogorelov$^1$, H. G. Silva$^2$, and J. F. Polido$^3$}
\affiliation{$^1$IFIMUP and IN-Institute of Nanoscience and Nanotechnology, Universidade
do Porto, Rua do Campo Alegre $687$, $4169-007$ Porto, Portugal, $^2$Geophysical Centre
of Évora and Physics Department, ECT, University of Évora, Rua Romão Ramalho 59, 7002-554
Évora, Portugal, $^3$Ecole Polytechnique F\'{e}d\'{e}rale de Lausanne, Station 1, 1015
Lausanne, Switzerland.}

\begin{abstract}
Tunnel transport processes are considered in a square lattice of metallic nanogranules
embedded into insulating host to model tunnel conduction in real metal/insulator granular
layers. Based on a simple model with three possible charging states ($\pm$, or $0$) of a
granule and three kinetic processes (creation or recombination of a $\pm$ pair, and charge
transfer) between neighbor granules, the mean-field kinetic theory is developed. It
describes the interplay between charging energy and temperature and between the applied
electric field and the Coulomb fields by the non-compensated charge density. The resulting
charge and current distributions are found to be essentially different in the free area
(FA), between the metallic contacts, or in the contact areas (CA), beneath those contacts.
Thus, the steady state dc transport is only compatible with zero charge density and ohmic
resistivity in FA, but charge accumulation and non-ohmic behavior are \emph{necessary} for
conduction over CA. The approximate analytic solutions are obtained for characteristic
regimes (low or high charge density) of such conduction. The comparison is done with the
measurement data on tunnel transport in related experimental systems.
\end{abstract}

\maketitle\

\section{\lb{int}Introduction}

More than three decades have passed since the pioneering studies by Abeles and co-workers
\cite{sheng1,sheng2} that triggered a huge research effort in granular thin films. Actually,
nanostructured granular films are of a considerable interest for modern technology due to
their peculiar physical properties, like giant magnetoresistance \cite{Berk}, Coulomb
blockade \cite{fert,varalda}, or high density magnetic memory \cite{park}, impossible for
continuous materials. 

However, a number of related physical mechanisms still needs better understanding, in 
particular, transport phenomena in these films are still a great challenge and presently 
various works are addressing such problem \cite{beloborodov,kozub,ng}. The main reason is 
that granular systems reveals certain characteristics which cannot be obtained neither in 
the classical conduction regime (in metallic, electrolyte, or gas discharge conduction) 
nor in the hopping regime (in doped semiconductors or in common tunnel junctions).
Their specifics is mainly determined by the drastic difference between the characteristic
time of an individual tunneling event ($\sim \hbar/ \e_{\rm F} \sim 10^{-15}$ s) and the
interval between such events on the same granule $\sim e/(j d^2) \sim 10^{-3}$ s, at typical
current density $j \sim 10^{-3}$ A/cm$^2$ and granule diameter $d \sim 5.0$ nm. Other
important moments are the sizeable Coulomb charging energy $E_c \sim e^2/(\e_{\rm eff}d)$
(typically $\sim 10$ meV) and the fact that the tunneling rates across the layer may be even
several orders of magnitude slower than along it. The interplay of all these factors leads to
unusual macroscopic effects, including a peculiar slow relaxation of electric charge
discovered in experiments on tunnel conduction through granular layers and granular films
\cite{Kak1,Sch}.

For theoretical description of transport processes in granular layers (and multilayers) we
develop an extension of the classical Sheng-Abeles model for a single layer of identical 
spherical particles located in sites of a simple square lattice, with three possible charging 
states ($\pm$, or $0$) of a granule and three kinetic processes: creation of a $\pm$ pair (the 
only process included in the original Sheng-Abeles treatment) on neighbor granules, recombination 
of such a pair, and charge translation from a charged to neighbor neutral granule. Even this 
rather simple model, neglecting the effects of disorder within a layer and of multilayered 
structure, reveals a variety of possible kinetic and thermodynamical regimes, well resembling 
those observed experimentally. 

The detailed formulation of the model, its basic parameters, and its mean-field continuum version 
are given in Sec. \ref{char}. Next in Sec. \ref{mf} we calculate the mean values of occupation 
numbers of each charging state under steady state conditions, including the simplest equilibrium 
situation (no applied fields), in function of temperature. The analysis of current density and 
related kinetic equation in the out-of-equilibrium case is developed in Sec. \ref{fa}, where also 
its simple, ohmic solution is discussed for the FA part of the system. The most non-trivial regimes 
are found for the CA part, as described in Sec. \ref{ca} for steady state conduction with charge 
accumulation and non-ohmic behavior. The general integration scheme for non-linear differential 
equation, corresponding to steady states in FA and CA, and particular approximations leading to 
their analytic solutions are dropped into Appendix.

\section{\lb{char}Charging states and kinetic processes}

\begin{figure}
  \includegraphics[width=8cm]{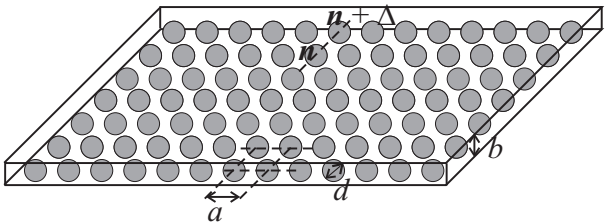}\\
  \caption{Square lattice of metallic granules in the insulating matrix.}
  \label{fig1}
\end{figure}

We consider a system of identical spherical metallic nanogranules of diameter $d$, located
in sites of simple square lattice of period $a$ within a layer of thickness $b \sim a$ of
insulating host with a dielectric constant $\e$ (Fig. \ref{fig1}).
In the charge transfer processes, each granule can bear different numbers $\s$ of electrons
in excess (or deficit) to the constant number of positive ions and the resulting excess
charge $\s e$ defines a Coulomb charging energy $\sim \s^2 E_c$. At not too high temperatures,
$k_{\rm B}T \lesssim E_c$, the consideration can be limited only to the ground neutral state
$\s = 0$ and single charged states $\s = \pm 1$. Actually, for low metal contents (well
separated, small grains), $E_c$ reaches $\sim 10-30$ meV, so this approach can be reasonable
even above room temperature. For a three-dimensional ($3$D) granular array, $E_c$ was defined
in the classic paper by Sheng and Abeles \cite{sheng1}, under the assumption of a constant
ratio between the mean spacing $s$ and granule diameter $d$, in the form $E_c = e^2 f(s/d)/
(\e d)$, where the dimensionless function $f(z) = 1/(1 + 1/2z)$. Otherwise, the complete
dielectric response of 3D insulating host with the dielectric constant $\e$ and metallic
particles with the volume fraction $f < 1$ and diverging dielectric constant $\e_{m} \to \infty$
can be characterized by the effective value $\e_{eff} = \e/(1 - f)$.

For the planar lattice of granules, the analogous effective constant can be estimated, summing
the own energy $e^{2}/(\e d)$ of a charged granule at the $\bn = 0$ site and the energy of its
interaction with electric dipolar moments $\approx (e/\e_{eff})(d/2n)^{3}\bn$, induced
by the Coulomb field from this charge (in macroscopic dielectric approximation) on all the
granules at the sites $\bn = a(n_1,n_2)$:
\be
 E_{c} = \frac{e^{2}}{d}\left[\frac 1 \e - \frac \a {\e_{eff}^2} \left(\frac d a \right)^4
  \right] = \frac{e^2}{\e_{eff}d}.
   \lb{eq1}
    \ee
Here the constant $\a = \frac \pi 4 \sum_{n \ne 0} n^{-4} \approx 5.78$, and the resulting
$\e_{eff} = \left[\e + \sqrt{\e^2 + \e\a(d/a)^4}\right]/2 > \e$. However, Eq. \ref{eq1} may
considerably underestimate the most important screening from nearest neighbor granules at $d
\sim a$, and in what follows we generally characterize the composite of insulating matrix and
metallic granules by a certain $\e_{eff} = e^2/dE_c \gg \e$.

\begin{figure}[h!]
  \includegraphics[width=8cm]{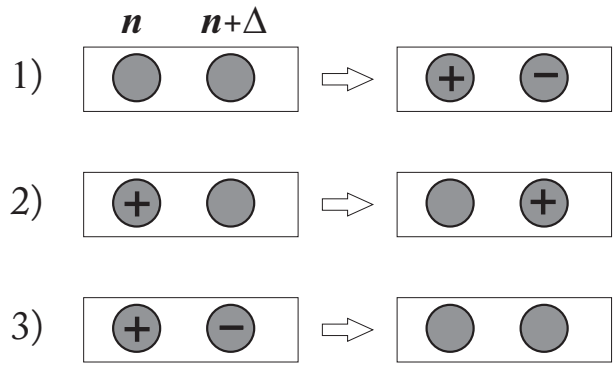}\\
  \caption{Kinetic processes in a granular layer.}\label{fig2}
\end{figure}

Following the approach proposed earlier \cite{Kak1}, we classify the microscopic states of
our system, attributing the charging variable $\s_\bn$ with values $\pm 1$ or $0$ to each
site $\bn$ and then considering three types of kinetic processes between two neighbor
granules $\bn$ and $\bn + \D$ (Fig. \ref{fig2}):

\begin{enumerate}
  \item \emph{Electron hopping from neutral $\bn$ to neutral $\bn + \D$}, creating a pair
  of oppositely charged granules: $(\s_\bn = 0,\s_{\bn + \D} = 0)\to (\s_\bn = +1, \s_{\bn
  + \D} = -1)$, only this process was included in the Sheng and Abeles' theory;
  \item \emph{Hopping of an extra electron or hole from $\bn$ to neutral $\bn + \D$}, that
  is the charge transfer: $(\s_\bn = \pm1,\s_{\bn + \D} = 0)\to (\s_\bn = 0,\s_{\bn + \D}
  = \pm1)$;
  \item \emph{Recombination of a electron-hole pair}, the inverse to the process 1.:
  $(\s_\bn = +1, \s_{\bn +   \D} = -1)\to (\s_\bn =0,\s_{\bn + \D} = 0)$.
\end{enumerate}

\begin{figure}[h!]
  \includegraphics[width=8cm]{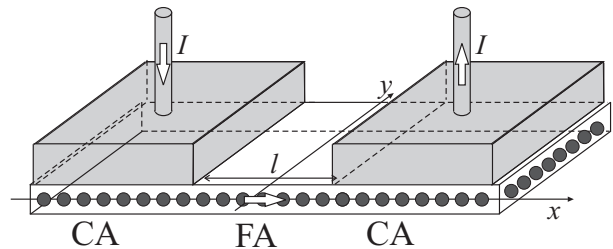}\\
  \caption{CIP conduction geometry.}\label{fig3}
\end{figure}

Note that all the processes 1) to 3) are conserving the total system charge $Q = \sum_{\bn}
\s_\bn$, hence the possibility for charge accumulation or relaxation only appears due to
the current leads. A typical configuration for current-in-plane (CIP) tunneling conduction
includes two macroscopic metallic electrodes on top of the granular layer, forming contact
areas (CA) where the current is being distributed from the electrodes into granules, through
an insulating spacer of thickness $b^{\prime}$, and a free area (FA) where the current
propagates over the distance $l$ between the contacts (Fig. \ref{fig3}). To begin with, we
consider a simpler case of FA while the specific analysis for CA with an account for screening
effects by metallic contacts will be given later in Sec. \ref{ca}.

The respective transition rates $q_{\bn,\D}^{(i)}$ for \emph{i}th process are determined by
the instantaneous charging states of two relevant granules and by the local electric field
$\mathbf{F}_\bn$ and temperature $T$, accordingly to the expressions:
\bea
 q_{\bn,\D}^{(1)} & = & \left(1 - \s_{\bn}^2 \right)\left(1 - \s_{\bn + \D}^2\right)
  \varphi\left(e{\mathbf F}_\bn\cdot\D + E_c\right)\nn \\
q_{\bn,\D}^{(2)} & = & \s_\bn^2 \left(1-\s_{\bn + \D}^2\right)\varphi
 \left(-e\s_{\bn}{\bf F}_\bn\cdot\D\right)\nn \\
q_{\bn,\D}^{(3)} & = & \frac{1}{2}\s_\bn \s_{\bn + \D}\left(\s_\bn \s_{\bn + \D}-1\right)
 \times \nn\\
& & \quad\quad \times\,\varphi\left(e\s_{\bn + \D}{\bf F}_\bn \cdot \D - E_c\right).
  \lb{eq2}
   \eea
Thus the charging energy is positive, $E_c$, for the pair creation, zero for the transport,
and negative, $-E_{c}$, for the recombination processes. The function $\varphi(E) = \o
N_{\rm F}E/[\exp(\b E)-1]$ expresses the total probability, at given inverse temperature
$\b = 1/(k_{\rm B}T)$, for electron transition between granules with Fermi density of states
$N_{\rm F}$ and Fermi levels differing by $E$. The hopping frequency $\o = \o_{a}\exp(-2\chi
s)$ involves the \emph{attempt frequency}, $\o_{a} \sim E_{\rm F}/\hbar$, the inverse
tunneling length $\chi$ (typically $\sim 10$ nm$^{-1}$), and the inter-granule spacing $s =
a - d$. Local electric field ${\bf F}_\bn$ on $\bn$th site consists of the external applied
field ${\bf A}$ (site independent) and the Coulomb field ${\bf C}_\bn$ due to all other
charges in the system:
\be
 {\bf C}_\bn = \frac e{\e_{eff}} \sum_{\bn' \neq \bn}  \s_{\bn'}\frac{\bn' - \bn}{\left|\bn'
  - \bn\right|^3}.
   \lb{eq3}
    \ee
A suitable approximation is achieved with passing from discrete-valued functions $\s_\bn$ of
discrete argument $\bn = a(n_1, n_2)$ to their continuous-valued mean-field (MF) equivalents
$\s_\br = \left\langle \s_\bn\right\rangle_\br$ (mean charge density) and $\r_\br =
\left\langle \s_\bn^2 \right\rangle_\br$ (mean charge carrier density). These densities are
obtained by averaging over a wide enough area (that is, great compared to the lattice period
but small compared to the size of entire system or its parts) around \emph{any} point $\br$
in the plane (for simplicity, we drop the position index at averages $\langle\,\rangle_\br$
in what follows). This also implies passing to a smooth local field:
\be
 {\bf F}_\br = {\bf A} + \frac e{\e_{eff}a^2} \int\s(\br') \frac{\br' - \br}{|\br' - \br|^3}
  d\br'.
   \lb{eq4}
    \ee
and to the averaged transition rates $q^{(i)}_{\br,\D} = \left\langle q_{\bn,\D}^{(i)}
\right\rangle$ and $p^{(i)}_{\br,\D} = \left\langle \s_\bn q_{\bn,\D}^{(i)}\right\rangle$.
These rates fully define the temporal derivatives of mean densities:
\bea
 \dot\s_\br & = & \sum_\D \left[q^{(1)}_{\br,\D} - q^{(1)}_{\br + \D, -\D} - p^{(2)}_{\br,\D}
  + p^{(2)}_{\br + \D,-\D}\right.\nn\\
 & & \qquad\qquad\qquad\qquad\qquad\qquad\qquad - \left. p^{(3)}_{\br,\D}\right],\lb{eq5}\\
  \dot\r_\br & = & \sum_\D \left[q^{(1)}_{\br,\D} + q^{(1)}_{\br + \D, -\D} - q^{(2)}_{\br,\D}
   + q^{(2)}_{\br + \D,-\D}\right.\nn\\
 & & \qquad\qquad\qquad\qquad\qquad\qquad\qquad - \left.q^{(3)}_{\br,\D}\right].
     \lb{eq6}
      \eea
The set of Eqs. \ref{eq2}-\ref{eq6} provides a continuous description of the considered
system, once a proper averaging procedure is established.

\section{Mean-field densities in equilibrium}
\lb{mf}
We perform the above defined averages in the simplest assumption of no correlations between
different sites: $\left\langle f_\bn g_{\bn^\prime}\right\rangle = \left\langle f_\bn
\right\rangle\left\langle g_{\bn^\prime}\right\rangle $, $\bn^\prime\neq\bn$, and using the
evident rules:
$\left\langle \s_\bn^{2k+1}\right\rangle = \s_\br$, $\left\langle \s_\bn^{2k}\right\rangle
= \r_\br$.
The resulting averaged rates are:
\bea
  q^{(1)}_{\br,\D} & = & \s^0_\br\s^0_{\br + \D}\varphi\left(e{\bf F}_\br\cdot\D +
  E_c\right),\nn\\
q^{(2)}_{\br,\D} & = & \s^0_{\br + \D}\left[\s^+_\br\varphi\left(-e{\bf F}_\br\cdot\D\right)
 \right.\nn\\
& & \qquad\qquad\qquad + \left.\s^{-}_\br\varphi\left(e{\bf F}_\br\cdot\D\right)\right],\nn\\
p^{(2)}_{\br,\D} & = & \s^0_{\br + \D}\left[\s^+_\br\varphi\left(-e{\bf F}_\br\cdot\D\right)
  \right.\nn\\
& & \qquad\qquad\qquad - \left.\s^-_\br\varphi\left(e{\bf F}_\br\cdot\D\right)\right],\nn\\
q^{(3)}_{\br,\D} & = & \left[\s^+_\br\s^-_{\br + \D}\varphi\left(-e{\bf F}_\br\cdot\D -
 E_c\right)\right.\nn\\
& & \qquad + \left.\s^-_\br\s^+_{\br + \D}\varphi\left(e{\bf F}_\br\cdot\D - E_c\right)
  \right],\nn\\
p^{(3)}_{\br,\D} & = & \left[\s^+_\br\s^-_{\br + \D}\varphi\left(-e{\bf F}_\br\cdot\D -
 E_c\right)\right.\nn\\
& & \qquad\qquad - \left.\s^-_\br \s^+_{\br + \D}\varphi\left(e{\bf F}_\br\cdot\D -
 E_c\right)\right],
  \lb{eq7}
   \eea
where the mean occupation numbers for each charging state $\s^\pm_\br = (\r_\br \pm \s_\br)/2$
and $\s^0_\br = 1 - \r_\br$ satisfy the normalization condition: $\sum_{i}\s^i_\br = 1$.

In a similar way to Eq. \ref{eq5}, we express the vector of average current density
${\bf j}_\bn$ at $\bn$th site:
\bea
 {\bf j}_\bn & = & \frac e{a^2 b}\sum_\D \D\left[-q^{(1)}_{\bn,\D} + q^{(1)}_{\bn + \D,-\D}
  \right. \nn\\
& + & \left. p^{(2)}_{\bn,\D} - p^{(2)}_{\bn + \D,-\D} + p^{(3)}_{\bn,\D}\right],
  \lb{eq8}
   \eea
and then its MF extension $\bf{j}_\br$ is obtained by simple replacing $\bn$ by $\br$ in
the arguments of $q^{(i)}$ and $p^{(i)}$. Expanding these continuous functions in powers
of $|\D| = a$, we conclude that Eq. \ref{eq5} gets reduced to usual continuity equation:
\be
 \dot{\s}_\br = - \frac{a^2b}e \nabla_2\cdot \bf{j}_\br,
  \lb{eq9}
   \ee
with the two dimensional ($2$D) nabla: $\nabla_2 = (\pd_x,\pd_y)$. We begin the analysis of
Eqs. \ref{eq5} - \ref{eq9} from the simplest situation of thermal equilibrium in absence of
electric field, $\bf{F_r} \equiv 0$, then Eq. \ref{eq5} turns into evident identity:
$\s_\br \equiv 0$, that means zero charge density, and Eq. \ref{eq8} yields in zero current
density: ${\bf j}_\br \equiv 0$, while Eq. \ref{eq6} provides a finite and constant value
of charge carrier density:
\be
 \r_\br \equiv \r_e = \frac 2{2 + \exp\left(\b E_c/2\right)}.
  \lb{eq10}
   \ee
At low temperatures, $\b E_{c} \gg 1$, this value is exponentially small: $\r_e \approx 2
\exp(-\b E_{c}/2)$, and for high temperatures, $\b E_c \ll 1$, it behaves as $\r_e \approx
\r_{\infty} - \b E_c/9$, tending to the limit $\r_{\infty} = 2/3$, corresponding to
equipartition between all three fractions $\s^i$ (Fig. \ref{fig4}, though this limit being
beyond actual validity of the model, as indicated in Sec. \ref{char}).
\begin{figure}
  \includegraphics[width=9cm]{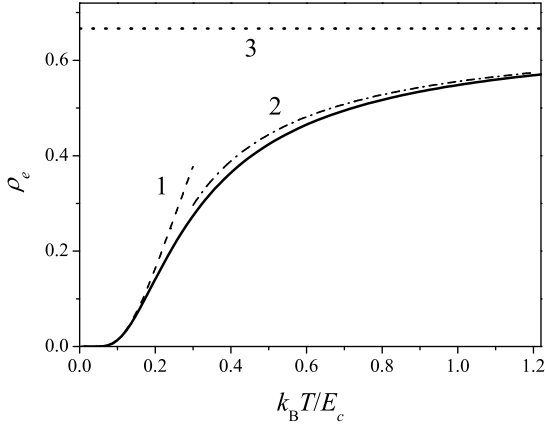}\\
  \caption{Equilibrium density $\r_e$ of charge carriers in function of temperature (solid
  line).   The curve $1$ (dashed line) corresponds to the low temperature asymptotics $\r_e
  \approx 2 \exp \left(-E_c/2k_{\rm B}T\right)$, and the curve 2 (dash-dotted line) to the high
  temperature asymptotic $\r_e \approx \r_{\infty} - E_c/9 k_{\rm B}T$, converging to the limit
  $\r_{\infty} = 2/3$ (dotted line).}
  \lb{fig4}
\end{figure}

In presence of electric fields ${\bf F}_\br \neq 0$, the local equilibrium should be
perturbed and the system should generate current and generally accumulate charge. Then, from
Eq. \ref{eq6}, the charge density $\s_\br$ is related to the carrier density $\r_\br$ as:
\be
 \s^2_\br = \frac{\left(\r_\br - \r_e\right)\left(\r_\br + \r_e - 2\r_e\r_\br\right)}
  {\left(1 - \r_e\right)^2},
   \lb{eq11}
    \ee
describing the increase of charge density with going away from equilibrium. As seen from Fig.
\ref{fig5}, for not too high temperatures $T \lesssim E_c/k_{\rm B}$ where the neglect of
multiple charged states is justified, this dependence is reasonably close to the simplest
low-temperature form:
\be
 \s \approx \sqrt{\r^2 - \r_e^2},
  \lb{eq12}
   \ee
that will be practically used in what follows.

Now we are in position to pass to the out-of-equilibrium situations, beginning from a
simpler case of dc current flowing through the FA.
\begin{figure}
\center \includegraphics[width=9cm]{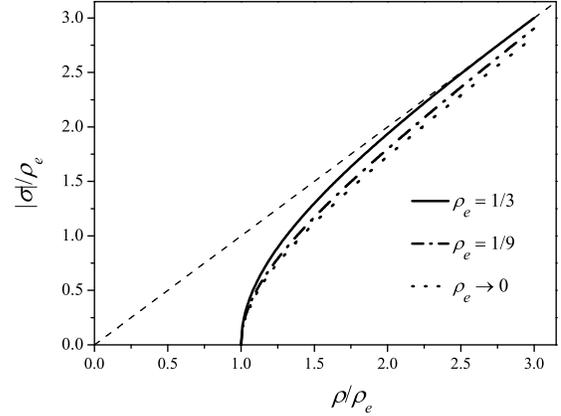}\\
\caption{The charge density $\s$ in function of the carrier density $\r$ for different
temperatures (corresponding to different thermal equilibrium values $\r_e$). Note closeness
of all the curves to that for low-temperature limit $\r_e \to 0$, given exactly by Eq.
\ref{eq12}.}
\lb{fig5}
\end{figure}

\section{\label{fa}Steady state conduction in FA}

In presence of (generally non-uniform) fields ${\bf F}_\br$ and densities $\s_\br$, $\r_\br$,
we expand Eq. \ref{eq8} up to 1st order terms in $|\D| = a$ and obtain the local current
density as a sum of two contributions, the field-driven and diffusive:
\be
 {\bf j}_\br = {\bf j}^{\rm field}_\br +  {\bf j}^{\rm dif}_\br = g\left(\r_\br \right)
  {\bf F}_\br - eD \left(\r_\br\right)\nabla_2\s_\br,
   \lb{eq13}
    \ee
where the effective conductivity $g$ and diffusion coefficient $D$ are functions of the local
charge carrier density, $\r \equiv \r_\br$:

\bea
 g(\r) & = & \frac{e^2}b \left|\frac{}{}2(1 - \r)^2\varphi'\left(E_c\right) + \r(1 - \r)
 \varphi'(0)\right.\nn\\
& + & \left.\frac 12 (\r^2 - \s^2) \varphi'\left(-E_c\right)\right|,\nn\\
 D(\r) & = & \frac{\r (1-\r_e)^2\varphi(0)(1 - \r)\r_e^2 \varphi(-E_c)/2}{\r(1 - 2\r_e) +
  \r_e^2}.
   \lb{eq14}
    \eea
In view of Eqs. \ref{eq11}, \ref{eq12}, we can consider $g$ and $D$ as \emph{even} functions of
local charge density $\s$, and just this dependence will be mostly used below. Also $g$ and $D$
depend on temperature, through the functions $\varphi$ and $\varphi'$. The system of Eqs.
\ref{eq11} -\ref{eq14}, together with Eq. \ref{eq4}, is closed and self-consistent, defining the
distributions of $\s_\br$ and $\r_\br$ at given ${\bf j}_\br$. It is readily seen to admit the
trivial solution, $\s(x) \equiv 0$, and now we shall argue that in fact this is the only practical
solution for FA.

First of all, we notice physical restrictions on the charge accumulation in FA. By the problem
symmetry, the charge density should only depend on the coordinate along the current, $\s =
\s(x)$, this function being odd (in the geometry of Fig. \ref{fig3}) and supposedly monotonous.
Then its maximum value $\s_{\rm max} = \s(L/2)$ will define the characteristic scale for the
Coulomb field: $C \sim \s_{\rm max}e/\left(\e_{\rm eff}a^2\right)$ which should not be higher
than typical applied fields $A \sim 10^2$ V/cm (as seen from relatively moderate non-ohmic vs
ohmic response in the experiment). Thus the maximum charge density should not surpass the level
of $A \e_{\rm eff}a^2/e \sim 10^{-3}$, that is much lower than the equilibrium density of charge
carriers $\r_e$ (except for, maybe, too low temperatures, $T \lesssim 0.07 E_c/k_{\rm B} \sim
10$ K). Therefore, one can neglect the small difference, Eq. \ref{eq12}, setting constant
values: $\r \approx \r_e$ and then $g \approx g_e \equiv g(\r_e),\, D \approx D_e \equiv
D(\r_e)$.

Under such condition, we can eliminate the (not well known) constant $A$ from Eq. \ref{eq13},
bringing this equation to the integro-differential form:
\be
 \frac{\pd^2 \s(x)}{\pd x^2} = \frac{g_e}{D_e \e_{\rm eff} a^2} P \int_{-l/2}^{l/2}
  \frac{\s(x')dx'}{(x - x')^2},
   \lb{eq15}
    \ee
where the $P$-symbol at integration in $x'$ means the "discrete principal value", that is
omission of the interval $(x - a, x + a)$ to avoid the apparent divergence, in agreement
with the minimum distance between granules in the lattice. Thus the regularized integral
converges rapidly, then it is reasonable to fix the argument of $\s$-density at $x' = x$,
arriving at a simple differential equation:
\be
 \frac{\pd^2 \s(x)}{\pd x^2} = \frac{\s(x)}{r_\b^2}.
  \lb{eq16}
   \ee
Here the parameter
\[r_\b^2 = \frac{a^3}d \frac{2{\rm e}^{\b E_c} + 5{\rm e}^{\b E_c/2}+2}{{\rm e}^{3\b E_c/2}
+ 2\b E_c{\rm e}^{\b E_c} - {\rm e}^{\b E_c/2}}\]
defines the temperature dependent length scale $r_\b$, and the $x$-odd solution of Eq.
\ref{eq16} is just $\s(x) = \s_1 \sinh(x/r_\b)$. However, for all the considered temperatures,
$\b E_c \gtrsim 1$ (see the note in Sec. \ref{char}), this scale is $r_\b \lesssim a$, that is
by many orders of magnitude smaller than the FA size $l$. Then the estimate for the constant
$\s_1$ in the above solution, $\s_1 \sim \s_{\rm max}{\rm e}^{-l/r_\b}$ with the exponent as
great as for instance $l/r_\b \sim 10^4$, makes this solution practically vanishing within
whole FA, except maybe for a very narrow vicinity $\sim r_\b$ of its interface with CA (where,
strictly speaking, Eq. \ref{eq16} no more holds). This evident consequence of long-range
character of Coulomb fields in FA will be contrasted below with the situation in CA, where
charge accumulation turns possible due to screening effects by the metallic contacts and to
the related short-range fields.

Thus we conclude that there is practically no charge accumulation and hence no diffusive
contribution to the current in FA. Thus the steady state of FA in out-of-equilibrium
conditions should be characterized by the ohmic conductivity $g_e$. In fact, an estimation
(based on an experimental system \cite{silva}) suggests that the FA contribution to the
overall resistance turns to be about two orders of magnitude smaller than the CA one (see
below), and thus the transport is expected to be mainly controlled by CA.

\section{\label{ca}Steady state conduction in CA}

\begin{figure}
  \includegraphics[width=8cm]{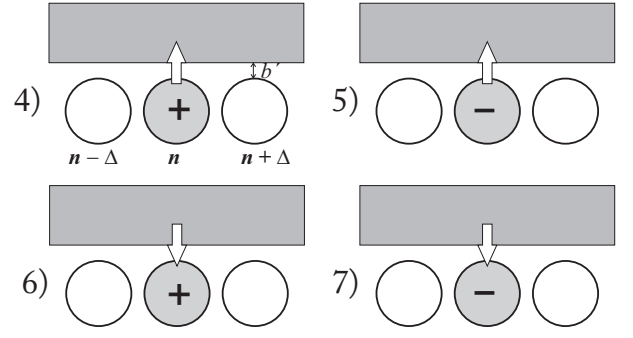}\\
   \caption{Kinetic processes between $\bn$th granule and the metallic electrode in CA.}
    \lb{fig6}
\end{figure}

The kinetics in CA includes, besides the processes 1) to 3) of Sec. \ref{mf} and \ref{fa},
also four additional microscopic processes between $\bn$th granule and the electrode (Fig.
\ref{fig6}) which are just responsible for variations of total charge $Q$ by $\pm 1$. The
respective rates $q^{(i)}$, $i = 4,\ldots 7$, are also dependent on the charging state
$(\s_\br,\r_\br)$ of the relevant granule and, using the same techniques that before, their
mean values are:
\bea
 q^{(4)}_\br& = & \left(\r_\br + \s_\br\right)\psi(-U - E_{c}'),\nn \\
 q^{(5)}_\br& = &\left(\r_\br - \s_\br\right)\psi\left(U - E_{c}'\right),\nn \\
 q^{(6)}_\br& = & \left(1 - \r_\br\right)\psi\left(U + E_{c}'\right),\nn \\
 q^{(7)}_\br& = &\left(1 - \r_\br\right)\psi\left(-U + E_{c}'\right).
  \lb{eq17}
   \eea
Here the function $\psi(E)$ formally differs from $\varphi(E)$ only by changing the pre-factor:
$\o \to \o' = \o_{a}{\rm e}^{-2\chi b'} \ll \o$, but the arguments of these functions in Eq.
\ref{eq17} include other characteristic energies. Thus, the energy $U = eb'S$ is due to the
electric field $S \equiv F_c (z = b')$ at the contact surface above the granule. As seen from
Fig. \ref{fig7}, this field is always normal to the surface and its value is defined by the
local charge density $\s$ (see below). At least, the charging energy $E_{c}'$ for a granule
under the contact can be somewhat lower (e.g., by $\sim 1/2$) than $E_c$. Then the kinetic
equations in interface region present a generalization of Eqs. \ref{eq5}-\ref{eq6}, as follows:
\bea
 \dot{\s}_\br& = & \sum_{{\D}}\left[q^{(1)}_{\br,\D} - q^{(1)}_{\br + \D,-\D}
  -p^{(2)}_{\br,\D}+ p^{\left(2\right)}_{\br + \D,-\D} - p^{(3)}_{\br,\D} \right. \nn \\
 & & \left.\quad\quad\quad  - q^{(4)}_\br + q^{(5)}_\br + q^{(6)}_\br - q^{(7)}_\br\right],
\lb{eq18}
  \eea
\bea
  \dot{\r}_\br & = & \sum_\D\left[q^{(1)}_{\br,\D} + q^{(1)}_{\br + \D,-\D}
  -q^{(2)}_{\br,\D} + q^{(2)}_{\br + \D,-\D} - q^{(3)}_{\br,\D}\right. \nn \\
 & &\left.\quad\quad\quad  -q^{(4)}_\br - q^{(5)}_\br + q^{(6)}_\br + q^{(7)}_\br\right].
  \lb{eq19}
   \eea
\begin{figure}
  \includegraphics[width=8cm]{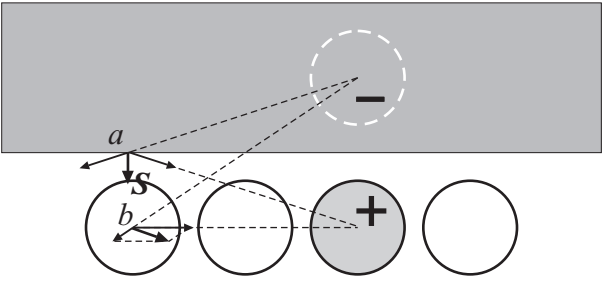}
  \caption{Formation of local electrical fields by a dipole of a charged granule and its
  (oppositely charged) image: at the surface of the metallic electrode (point \emph{a}) and
  on other granule (point \emph{b}).}
  \lb{fig7}
\end{figure}

The additional terms, by the \emph{normal} processes 4) to 7), are responsible for
appearance of a \emph{normal} component of current density:
\be
 j^z_\br = \frac{e}{a^{2}}\left[q^{(4)}_\br - q^{(5)}_\br - q^{(6)}_\br + q^{(7)}_\br\right],
  \lb{eq20}
   \ee
besides the planar component, still given by Eq. \ref{eq8}. But an even more important
difference from the FA case is the fact that the Coulomb field here is formed by a
\emph{double layer} of charges, those by granules themselves and by their images in the
metallic electrode (Fig. \ref{fig7}). Summing the contributions from all the charged granules
and their images (except for the image of $\bn$th granule itself, already included in the
energy $E_{c}'$), we find that the above mentioned field at the contact surface above the
point $\br$ of the granular layer, $S_\br$, can be expressed as a \emph{local} function of
the charge density $\s_\br$:
\be
  S_\br = C_\br(z = b^\prime) = - \frac{4\pi e}{\e a^{2}}\s_\br,
  \lb{eq21}
   \ee
replacing the integral relations, Eqs. \ref{eq3}-\ref{eq4}, in FA. Also, note that the
relevant dielectric constant for this field formed outside the granular layer is rather the
host value $\e$ than the renormalized $\e_{eff}$ within the layer (as by Eq. \ref{eq3}).
Then, the planar component of the field by charged granules ${\bf F }^{pl}_\br =
{\bf C}_\br(z = 0)$ is determined by the above defined normal field $S_\br$ through the
relation ${\bf F}^{pl}_\br = b' \nabla_{2}S_\br$. The density of planar current is
${\bf j}^{pl}_\br = g{\bf F}^{pl}_\br - e D{\nabla}_2\s_\br$, accordingly to Eq. \ref{eq13},
that is both field-driven and diffusive contributions into ${\bf j}^{pl}_\br$ are present
here and both they are proportional to the gradient of $\s_\br$. In the low temperature
limit, this proportionality is given by:
\be
 {\bf j}^{pl}_\br \approx - \left[\frac{8\pi e^3\o N_{\rm F} b'}{\e_{eff }a^3}
  g\left(\s_\br\right) + \frac{e\o N_{\rm F} k_{\rm B}T} a\right]\nabla_2\s_\br.
  \lb{eq22}
   \ee
Note that the presence of a non-linear function:
\[g(\s) = \sqrt{\r_{e}^{2} + \s^2}-2\r_{e}^{2}-\s^{2},\]
defines a \emph{non-ohmic} conduction in CA. In fact, this function should be defined by
Eq. \ref{eq21} only for charge density below its maximum possible value $|\s_{max}|=
\sqrt{1 - \r_{e}^2}$, turning zero for $|\s| > |\s_{max}|$ (note that the latter
restriction just corresponds to our initial limitation to the single charged states, see
Sec. \ref{char}). In the same limit of low temperatures, the normal current density is
obtained from Eqs. \ref{eq16}, \ref{eq17} as ${\bf j}_z(\br) = G_z\S_\br$ where $G_z \approx
\o' N_{\rm F} E'_c \e_{eff}/4\pi$. Finally, the kinetic equation in this case is obtained,
in analogy with Eq. \ref{eq8}, as:

\be
 \dot\s_\br = -\frac{a^2 b} e \nabla_2 \cdot {\bf j}^{pl}_\br + \frac{a^2}ej^z_\br.
  \lb{eq23}
   \ee

This equation permits to describe the steady state conduction as well as various time
dependent processes. The first important conclusion is that steady state conduction in
the interface turns only possible at non-zero charge density gradient, that is,
\emph{necessarily} involving charge accumulation, in contrast to the above considered
situation in bulk.

Let us restrict here the analysis to the steady state conduction regime which is simpler,
though the obtained results can be also used for the analysis of a more involved case when
an explicit temporal dependence of charge density is included in Eq. \ref{eq23} (this will
be a topic of future study).

\begin{figure}
  \includegraphics[width=8cm]{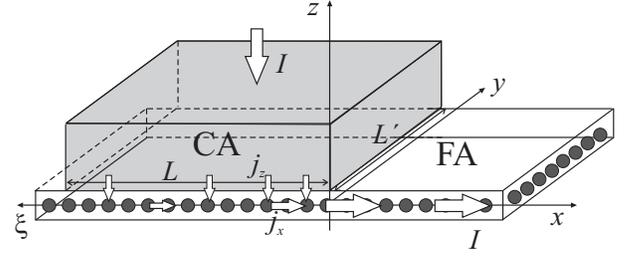}\\
  \caption{Relations between longitudinal ($j_x$) and normal ($j_z$) currents in CA, adding
  to the total current $I$.}
  \label{fig8}
\end{figure}

We choose the contacts geometry in the form of a rectangular stripe of planar dimensions
$L \times L^{\prime}$, along and across the current respectively. In neglect of relatively
small effects of current non-uniformity along the lateral boundaries, the only relevant
coordinate for the problem is longitudinal, $x$ (Fig. \ref{fig8}), so we consider the
relevant function $\s_x$ with its derivatives, spatial $\s_x'$ and temporal $\dot{\s_x}$.
In the steady state regime, $\dot{\s} = 0$ in Eq. \ref{eq23}, and the total current $I =
\rm{const}$, defined by the action of external source. Then, using the above approximation
for $g(\s)$, a non-linear 2nd order equation for charge density is found:
\be
 \frac{d}{dx}\left\{\left[g\left(\s_x\right) + \t\right]\s_x'\right\} - k^2\s_x
  = 0.
   \lb{eq24}
    \ee
Here the parameters are: $k^2 =(\o'E_{c}')/\left(ab\o k_{\rm B}T_1\right)$ and $\t = T/T_1$,
where $T$ is the actual temperature and $T_1 = 8\pi e^2 b'/a^2 k_{\rm B}\e_{eff}$. To define
completely its solution, the following boundary conditions are imposed:
\be
 \s_{x = 0}' = \frac{k^2 b'\s_{x = 0}}{g\left(\s_{x = 0}\right) + \t},
   \lb{eq25}
    \ee
\be
 \s_{x = L}' = \frac{a}{Le\o b N_{\rm F}k_{\rm B}T_1}\frac{I}{g\left(\s_{x=L}\right) + \t}.
   \lb{eq26}
    \ee

Here Eq. \ref{eq25} corresponds to the fact that the longitudinal current $j^x$ at the
initial point of contact/granular sample interface (the leftmost in Fig. \ref{fig8}) is
fully supplied by the normal current $j^z$ entering from the contact to the granular sample,
and Eq. \ref{eq26} corresponds to the current continuity at passage from CA (of length $L$
along the $x$ axis) to FA.

Let us discuss the solution of Eq. \ref{eq24} qualitatively. Generally, to fulfill the
conditions, Eqs. \ref{eq25}, \ref{eq26}, one needs a quite subtle balance to be maintained
between the charge density and its derivatives at both ends of contact interface. But the
situation is radically simplified when the length $L$ is much greater than the characteristic
decay length for charge and current density: $kL \gg 1$. In this case, the relevant coordinate
is $\xi = L-x$, so that the boundary condition \ref{eq25} corresponds to $\xi = L \to \infty$,
when both its left and right hand side turn zeros:
\be
 \s_{\xi\to\infty} = 0, \qquad \s_{\xi\to\infty}' = 0.
  \lb{eq27}
   \ee
The numeric solution shows that, for any initial (with respect to $\xi$, that is related to
$x = L$, Eq. \ref{eq26}) value of charge density $\s_{\xi = 0} = \s_0$, there is a
\emph{unique} initial value of its derivative $\s_{\xi = 0}' = D(\s_0)$ which just assures
the limits, Eq. \ref{eq27}, while for $\s_{\xi = 0}' > D(\s_0)$ the asymptotic value diverges
as $\s_{\xi\to\infty} \to \infty$, and for $\s_{\xi = 0}' < D(\s_0)$ it diverges as
$\s_{\xi\to\infty} \to -\infty$. Then, using the boundary condition, Eq. \ref{eq26}, and
taking into account the relation $V = V_0 \s_0$ following from Eq. \ref{eq23} with $V_0 =
4\pi e b'/(\e_{eff}a^2)$, we conclude that the function $D(\s_0)$ generates the
\emph{I}-\emph{V} characteristics:
\be
 I = I_{1}b' D\left(\frac{V}{V_0}\right)\left[g\left(\frac{V}{V_0}\right) + \t\right],
  \lb{eq28}
   \ee
where $I_{1} = e\o N_{\rm F}k_{\rm B}T_1$.

A more detailed analysis of Eq. \ref{eq24} is presented in Appendix. In particular, for
the weak current regime (Regime I) when $\s_0 \ll \s_1 = \sqrt{32\r_e(\r_e + \t)} \ll 1$,
so that $g(\s) \approx \r_e + \s^2/\left(2\r_e\right)$ along whole the contact interfaces,
Eq. \ref{eq23} admits an approximate analytic solution:

\be
 \s_\xi = \s_0{\rm e}^{-\l \xi}\left[1 + 6\left(\frac{\s_0}{\s_1}\right)^2\left(1 -
  {\rm e}^{-2\l\xi}\right)\right],
   \lb{eq29}
    \ee
with the exponential decay index $\l = k/\sqrt{\r_e + \t}$.

This results in the explicit \emph{I-V} characteristics for Regime I:
\be
 I = G_0 V \left[1 + \left(\frac V {V_1}\right)^2\right],
  \lb{eq30}
   \ee
for $V < V_1 = \s_1 V_0$, Eq. \ref{eq30} describes the initial ohmic CA conductance
(temperature $\t$ dependent):
\be
  G_0  = \frac{I_1 k b^\prime}{V_0}\sqrt{ \r_e \left(\t \right) + \t},
   \lb{eq31}
    \ee
which turns non-ohmic for $V \sim V_1$. But at so high voltages another conduction regime
already applies (called Regime II), where $\s_1 \ll \s_0 \ll 1$ and one has $g(\s) \approx
\s$ (see Eq. \ref{eq21}). Following the same reasoning as for the Regime I, we obtain a
non-linear \emph{I}-\emph{V} characteristics for Regime II:
\be
 I \approx \frac{I_1 k b'}{\sqrt{3 V_0^3}}\left(V + V_0\t\right)^{3/2}
  \lb{eq32}
   \ee
this law is weaker temperature dependent than Eq. \ref{eq30}, which is related to the fact
that the conductance in Regime II is mainly due to dynamical accumulation of charge and not
to thermic excitation of charge carriers. Interestingly a $I \propto V^{3/2}$ law was recently
found in experimental measurements \cite{silva}. Further, such non-linearity can be yet more
pronounced if multiple charging states are engaged, as may be the case in real granular
layers with a certain statistical distribution of granule sizes present.

At least, for even stronger currents, when already $\s_0 \sim 1$, the solutions of Eq.
\ref{eq24} can be obtained numerically, following the above discussed procedure of adjustment
of the derivative $D(\s_0)$ to a given $\s_0$. Such solutions have an asymptotic behavior of
the type: $I \propto V^{5/4}$.

A simple and important exact relation for the total accumulated charge $Q$ in CA is obtained
from the direct integration of Eq. \ref{eq24}:
\[Q = t I,\]
where the parameter $t = 1/\psi\left(-E_{c}'\right)$ should have a role of characteristic
relaxation time in non-stationary processes. Assuming its value $ t \sim 1$ s (comparable
with the experimental observations \cite{Kak1}), together with the above used values of $\o$
and $T_1$, we conclude that the characteristic length scale $\l^{-1}$ for solutions of Eq.
\ref{eq24} can reach up to $\sim 10^{3} a \sim 1 \m$m, which is a reasonable scale for a
charge distribution beneath the contacts.

\section{\lb{global} Global conduction in the system}

The conduction in the overall system results from matching of the above considered processes
in CA and FA. Thus, in order to evaluate the global resistance of this circuit in series
it is necessary to add the contributions of both areas to it. Recent measurements \cite{silva}
have shown notably non-linear I-V curves (already at low enough voltages), so, accordingly
to the above discussion, this indicates that the resistance should be dominated by CA. To
have a clear view on it, we can use the typical parameters for the granular film: $a \sim
5$ nm, $d \sim 4$ nm, $\chi \sim 10$ nm$^{-1}$, $b \sim 8$ nm, $b' \sim 2$ nm, $E_c\sim 10$
meV, $N_F \sim 1$ eV$^{-1}$ and take $\o$ as a (less known) fitting parameter. For the
considered rectangular CIP geometry we also use the experimental values \cite{silva} of
width $L' = 3$ mm and of distance between the contacts $l = 100$ $\mu$m.

\begin{figure}
\includegraphics[width=9cm]{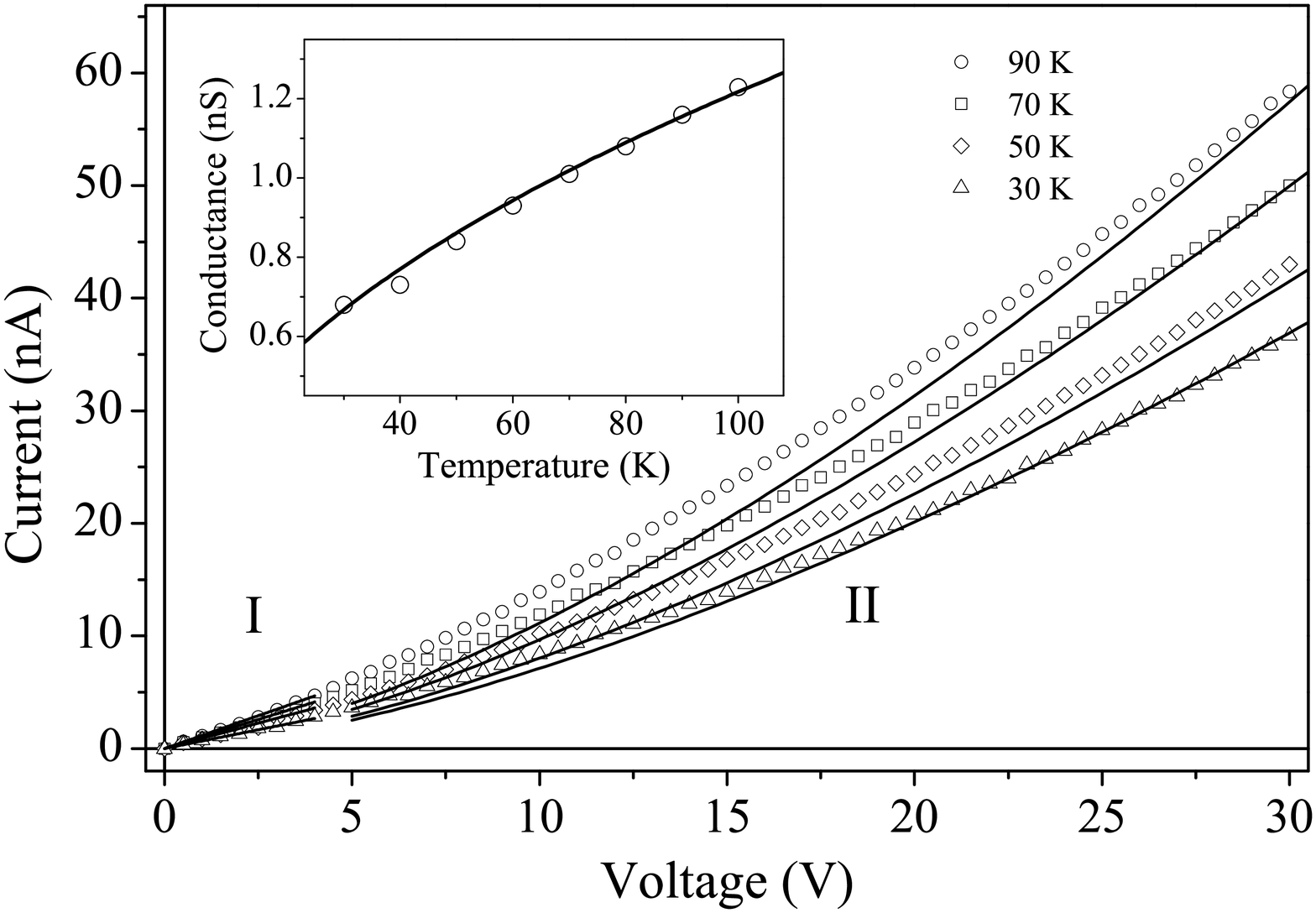}
\caption{I-V characteristics for a granular sample at different temperatures, compared
with the theoretical curves for for regimes I and II. Inset: temperature dependence of
ohmic conductance $G_0$, measured data (circles) vs calculated by Eq. \ref{eq31}.}
\label{fig9}
\end{figure}

Choosing $T = 50$ K, the ohmic conductance of the FA, $G_{FA}$, can be calculated through
the formula $G_{FA} = g(\r_e)L'b/l \approx \o \, 1.5\times10^{-18}$ S. In the CA, we can
estimate the conductance (in Regime I) following the above formula $G_{CA} \approx \o \,
8.0\times10^{-22}$ S. Thus it is clear that, for any choice of $\o$, the conductance of the
CA is about 4 orders of magnitude smaller than that of the FA and for that reason it should
dominate the global resistance of the system. Then, using the formulae, Eqs. \ref{eq29}-\ref{eq31},
we obtain a good agreement with the experimental data by Ref.
\cite{silva} as shown in Fig. \ref{fig9}. It should be noted however that the effective value
of the parameter $V_0$ giving the best fit to the experimental data should be notably higher
then that given by our formula (before Eq. \ref{eq28}) for single layer system. Thus, with
the above choice of other parameters, we have the single-layer value $V_0 \approx 0.5$ V whereas
the best fit for 10-layer experimental sample needs instead $V_{exp} \approx 3$ V. This difference
can be effectively accounted for by a simple multiplicative factor $\a \approx 6$ (the "multilayer
factor") so that $V_{exp} = \a V_0$ assures both the agreement for Regimes I,II of I-V curves
and the boundary $V \sim V_{exp}$ between them, clearly seen in Fig 9.

\section{Conclusion}
In conclusion, the mean-field model is developed for tunnel conduction in a granular layer,
including three principal processes of creation and annihilation of pairs of opposite charges
on neighbor granules and of charge transfer from a charged granule to a neighbor neutral granule.
Effective kinetic equations for averaged charge densities are derived for the characteristic
areas of the granular sample: the contact areas beneath metallic current leads and free area
between these leads. From these kinetic equations, it is shown that the tunnel conduction in the
free area does not produce any notable charge accumulation, and the conduction regime here is
purely ohmic. Contrariwise, such conduction in the contact area turns impossible without charge
accumulation, leading to generally non-ohmic conduction regime, since the contact area dominates
in the overall resistance. Approximate analytic treatment is developed for calculation of charge
density and tunnel current in two characteristic regimes: I) for weak charge accumulation (compared
to the thermal density of charge carriers) and II) for strong charge accumulation, leading to a
non-ohmic $I \propto V^{3/2}$ conduction law. The calculated I-V curves and temperature dependencies
are found in a good agreement with available experimental data. The proposed model can be further
developed for description of multilayer strucuture effects and also of non-stationary conduction
processes, like anomalous slow current relaxation \cite{Kak2}. Finally, the elastic effects of
Coulomb forces by charged granules can be included in order to explain the remarkable phenomenon
of resistive-capacitive switching \cite{silva2}, in granular layered conductors.

\section{Acknowledgements}
The authors are grateful to G.N. Kakazei, J.A.M. Santos, J.B. Sousa, J.P. Ara\'{u}jo, J.M.B. Lopes
dos Santos and H.L. Gomes for kind assistance and valuable help in various parts of this work.
One of us (HGS) gratefully acknowledges the support from Portuguese FCT through the grant SFRH/BPD/63880/2009.

\section{\label{ap}Appendix}

\renewcommand{\theequation}{A\arabic{equation}}

\setcounter{equation}{0}

Let us consider the equation:
\be
 \frac{d}{d\xi}\left[g(\s) + \t\right]\frac{d\s}{d\xi} - k^2\s = 0
  \lb{eqA1}
   \ee
with certain boundary conditions $\s(0) = \s_0$, $\s'(0) = \s_0'$, resulting from Eqs.
\ref{eq24}, \ref{eq25}. For a rather general function $g(\s)$ we can define the function
\be
 f(\s) = \int_0^\s g(\s')d\s',
  \lb{eqA2}
   \ee
then Eq. \ref{eqA1} presents itself as:
 \be
  \frac{d^2 F_\xi}{d\xi^2} = k^2 \s_\xi,
   \lb{eqA3}
    \ee
where $F_\xi \equiv f\left(\s_\xi\right) + \t\s_\xi$. Considered irrespectively of $\xi$:
\be
 f(\s) + \t \s = F,
  \lb{eqA4}
   \ee
this equation also defines $\s$ as a certain function of $F$: $\s = \s(F)$. Hence it is
possible to construct the following function:
\be
 \phi(F) = 2\int_0^F\s(F')dF'.
  \lb{eqA5}
   \ee
Now, multiplying Eq. \ref{eqA3} by $2dF/d\xi$, we arrive at the equation:
\be
 \frac{d}{d\xi}\left(\frac{dF}{d\xi}\right)^2 = k^2 \frac{d\phi}{d\xi},
  \lb{eqA6}
   \ee
with $\phi(\xi) \equiv \phi(F_\xi)$. Integrating Eq. \ref{eqA6} in $\xi$, we obtain a 1st
order separable equation for $F_\xi$:
\be
 \frac{dF}{d\xi} = \pm k\sqrt{\phi(F)}.
  \lb{eqA7}
   \ee

\begin{figure}
  \includegraphics[width=9cm]{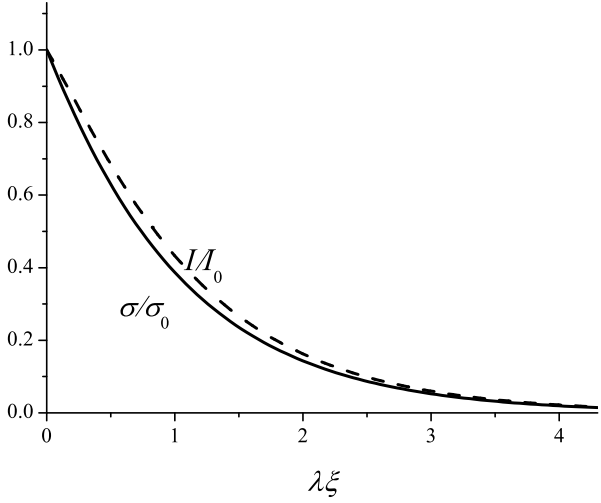}
  \caption{Charge density and current distribution in the CA region (Regime I).}\lb{fig12}
\end{figure}

We expect the function $F$ to decrease at going from $\xi = 0$ into depth of interface
region, hence choose the negative sign on r.h.s. of Eq. \ref{eqA7} and obtain its explicit
solution as:
\be
 \int_{F_\xi}^{F_0}\frac{dF'}{\sqrt{\phi(F')}} = k\xi
  \lb{eqA8}
   \ee
with $F_0 = f\left(\s_0\right) + \t \s_0$. Finally, the sought solution for $\s_\xi =
\s\left(F_\xi\right)$ results from substitution of the function $F_\xi$, given implicitly
by Eq. \ref{eqA8}, into $\s(F)$ defined by Eq. \ref{eqA4}. Consider some particular
realizations of the above scheme.
\begin{figure}
  \includegraphics[width=9cm]{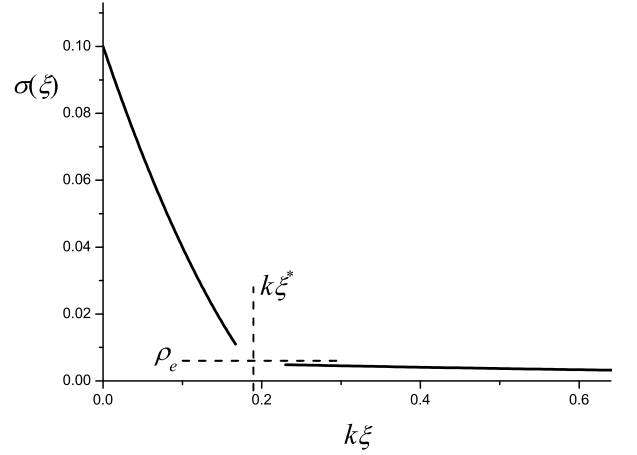}
  \caption{Charge density distribution in Regime II. A fast decay is changed to a slower
  exponential law,   after density dropping below the characteristic value $\r_e$.}
  \lb{fig13}
\end{figure}
For the approximate solution of $g(\s)$ given above, we have the explicit integral, Eq.
\ref{eqA2}, in the form:
\bea
  F\left(\s\right)& = & f(\s) + \t\s \nn \\
  & = &\left(\t + \frac{\sqrt{\r_e^2 + \s^2}}2 - \r_0^2 - \frac{\s^2}3 \right)\s + \nn\\
  & + &\r_e^2 \ln\sqrt{\frac{\s + \sqrt{\r_e^2 + \s^2}}{\r_e}}.
  \lb{eqA9}
   \eea
In the case $\s \ll \r_e \ll1$ (Regime I), Eq. \ref{eqA9} is approximated as:
\be
 F \approx \left(\r_e + \t\right)\s + \frac{\s^3}{6\r_e}
  \lb{eqA10}
   \ee
hence $\s(F)$ corresponds to a real root of the cubic equation, Eq. \ref{eqA10}, and in the
same approximation of Regime I it is given by:
\be
 \s(F) \approx \frac F{\r_e + \t}\left(1 - \frac{8F^2}{\s_1^2}\right),
  \lb{eqA11}
   \ee
with $\s_1 = 4\sqrt{\r_e\left(\r_e + \t\right)^3}$. Using this form in Eq. \ref{eqA5}, we
obtain:
\be
 \varphi(F) \approx \frac{F^2}{\r_e + \t}\left(1 - \frac{4F^2}{\s_1^2}\right),
  \lb{eqA12}
   \ee
and then substituting into Eq. \ref{eqA8}:
\be
 \ln\frac{\left[1 + \sqrt{1 - \left(2F/\s_1\right)^2}\right]F_0}{\left[1 + \sqrt{1 -
  \left(2F_0/\s_1\right)^2}\right]F} = \l\xi.
   \lb{eqA13}
    \ee
Inverting this relation, we define an explicit solution for $F_\xi$:
\be
 F(\xi) \approx F_0{\rm e}^{-\l\xi}\left[1 + \frac{F_0^2}{\s_1^2}\left(1 -
  {\rm e}^{-2\l\xi}\right)\right].
   \lb{eqA14}
    \ee
Finally, substituting Eq. \ref{eqA14} into Eq. \ref{eqA11}, we arrive at the result of Eq.
\ref{eq29} corresponding to Fig. \ref{fig12}.

For the regime II we have in a similar way:
\bea
  F(\s) & \approx & \s(\t + \s/2),\nn \\
  \s(F) & \approx & \sqrt{2F + \t^2} - \t, \nn \\
   \varphi(F) & \approx & \frac 3 2 \left[\left(2F + \t^2\right)^{3/2} - \t\left(3F +
    \t^2\right)\right] \nn \\
F_\xi & \approx & \left[F_0^{1/4} - \l_1\xi + \frac{3\t}{2^{5/4}\left(F_0^{1/4} -
 \l_1\xi\right)}\right]^4,
  \lb{eqA15}
   \eea
with $\l_1 = k/(2^{3/4}\sqrt{3})$, obtaining the charge density distribution (Fig. \ref{fig13}):
\be
\s(\xi) \approx \left(\sqrt{\s_0 + \t} - \l_1\xi\right)^2 - \t.
 \lb{eqA16}
  \ee

This function seems to turn zero already at $\xi = (\sqrt{\s_0 + \t} - \sqrt{\t})/\l_1$, but
in fact the fast parabolic decay by Eq. \ref{eqA16} only extends to $\xi \sim \xi^\ast$, such
that $\s_{\xi^\ast} \sim \r_e$, and for $\xi > \xi^\ast$ the decay turns exponential, like Eq.
\ref{eq29}. The \emph{I}-\emph{V} characteristics, Eq. \ref{eq32}, follows directly from Eq.
\ref{eqA16}.


\begin{thebibliography}{1}

\bibitem{sheng1} P. Sheng and B. Abeles, Phys. Rev. Lett. \textbf{28}, 34 (1972).

\bibitem{sheng2} P. Sheng, B. Abeles and Y. Aire, Phys. Rev. Lett. \textbf{31}, 44 (1973).

\bibitem{Berk} A.E. Berkowitz, J.R. Mitchell, M.J. Carey, A.P. Young, S. Zhang, F.E. Spada,
F.T. Parker, A. Hutten, G. Thomas, Phys. Rev. Lett. \textbf{68}, 3745 (1992).

\bibitem{fert} L.F. Schelp, A. Fert, F. Fettar, P. Holody, S.F. Lee, J.L. Maurice, F. Petroff,
A. Vaur\'{e}s, Phys. Rev. B \textbf{56}, R5747 (1997).

\bibitem {varalda} J. Varalda, W. A. Ortiz, A. J. A. Oliveira, B. Vodungbo, Y.-L. Zheng,
D. Demaille, M. Marangolo and D. H. Mosca, J. Appl. Phys. \textbf{101}, (2007) 014318.

\bibitem{park} M.A. Parker, K.R. Coffey, J.K. Howard, C.H. Tsang, R.E. Fontana, T.L. Hylton,
IEEE Trans. Magn. \textbf{32}, 142 (1996).


\bibitem{beloborodov} I. S. Beloborodov, A. V. Lopatin, V. M. Vinokur, and K. B. Efetov,
Rev. Mod. Phys. \textbf{79}, 469 (2007).

\bibitem{kozub} V. I. Kozub, V. M. Kozhevin, D. A. Yavsin, and S. A. Gurevich, JETP Lett.,
\textbf{81}, 226 (2005).

\bibitem{ng} Tai-Kai Ng and Ho-Yin Cheung, Phys. Rev B \textbf{70}, 172104 (2004).

\bibitem{Dieny}B. Dieny, S. Sankar, M.R. McCartney, D.J. Smith, P. Bayle-Guillemaud,
A.E. Berkowitz, J. Magn. Magn. Mater. \textbf{185}, 283 (1998).

\bibitem{Kak1}G.N. Kakazei, A.M.L. Lopes, Yu.G. Pogorelov, J.A.M. Santos, J.B. Sousa,
P.P. Freitas, S. Cardoso, E. Snoeck, J. Appl. Phys. \textbf{87}, 6328 (2000).

\bibitem{Sch}D. M. Schaadt, E.T. Yu, S. Sankar, A.E. Berkowitz, Appl. Phys. Lett. \textbf{74},
472 (1999).

\bibitem{Kak2} G. N. Kakazei, Yu.G. Pogorelov, A.M.L. Lopes, M.A.S. da Silva, J.A.M. Santos,
J.B. Sousa, S. Cardoso, P.P. Freitas, E. Snoeck, J. Magn. Magn. Mater. \textbf{266}, 62 (2003).

\bibitem{Kak3} G. N. Kakazei, P. P. Freitas, S. Cardoso, A. M. L. Lopes, Yu. G. Pogorelov,
J. A. M. Santos, J. B. Sousa, IEEE Trans. Mag. \textbf{35}, 2895 (1999).

\bibitem{lesnik} N. A. Lesnik, P. Panissod, G. N. Kakazei, Yu. G. Pogorelov, J. B. Sousa,
E. Snoeck, S. Cardoso, P. P. Freitas and P. E. Wigen, J. Magn. Magn. Mat. \textbf{485}, 242-245
(2002).

\bibitem{hazewinkel} M. Hazewinkel, Encyclopaedia of Mathematics, Kluwer Academic Publishers, 2001.

\bibitem{silva} H. G. Silva, H. L. Gomes, Y. G. Pogorelov, L. M. C. Pereira, G. N. Kakazei,
J. B. Sousa, J. P. Ara\'{u}jo, J. F. L. Mariano, S. Cardoso, and P. P. Freitas, J. Appl. Phys.
\textbf{106}, 113910 (2009).

\bibitem{silva2} H. Silva, H.L. Gomes, Yu.G. Pogorelov,  P. Stallinga, D.M. de Leeuw, J.P. Araujo,
J.B. Sousa, S.C.J. Meskers, G. Kakazei, S. Cardoso, P.P. Freitas, Appl. Phys. Lett. \textbf{94}, 202107,
2009.


\end{thebibliography}
\end{document}